\documentclass{elsart}

\usepackage{natbib}

\usepackage{epsfig}

\usepackage{amssymb}

\usepackage{subfigure}

\bibpunct{(}{)}{;}{a}{}{,} 

\begin{document}

\begin{frontmatter}



\title{Neutrino flares from black hole coronae}


\author{Florencia L. Vieyro} 
\address{Instituto Argentino de Radioastronom\'{\i}a (IAR, CCT La Plata, CONICET), C.C.5, (1894) Villa Elisa, Buenos Aires, Argentina }
\ead{fvieyro@iar-conicet.gov.ar}

\author{Gustavo E. Romero}
\address{Instituto Argentino de Radioastronom\'{\i}a (IAR, CCT La Plata, CONICET), C.C.5, (1894) Villa Elisa, Buenos Aires, Argentina \\ 
Facultad de Ciencias Astron\'omicas y Geof\'{\i}sicas, Universidad Nacional de La Plata, Paseo del Bosque s/n, 1900, La Plata, Argentina
}
\ead{romero@iar-conicet.gov.ar}


\begin{abstract}

We present a model for neutrino flares in accreting black holes based on the injection of a non-thermal population of relativistic particles in a magnetized corona. The most important products of hadronic and photohadronic interactions at high energies are pions. Charged pions decay into muons and neutrinos; muons also decay yielding neutrinos. Taking into account these effects, coupled transport equations are solved for all species of particles and the neutrino production is estimated for the case of accreting galactic black holes. 

\end{abstract}


\end{frontmatter}

\parindent=0.5 cm

\section{Introduction}

The existence of a broad-band X-ray/soft gamma-ray spectrum of accreting black holes in binary systems strongly suggests the presence of a very hot plasma (corona) around the central object. This plasma Comptonizes the soft X-ray photons produced in the inner accretion disk \citep{shakura} generating a power-law, high-energy feature in the spectrum. A wide variety of stable two-temperature corona models have been developed since \citet{shapiro} introduced a temperature differenciation for ions and electrons in the inner accretion flow of Cygnus X-1 (see, for instance, the review by \citealt{poutanen03}). 

Galactic black holes are found in different spectral states. The most typical are the \textsl{high-soft}, in which the spectrum is dominated by thermal emission of the disk, and the \textsl{low-hard}, characterized by the presence of a hot corona around the compact object and steady jets. The existence of these jets is supported by observational evidence of radio emission in almost every black hole binary in the hard state \citep{corbel2000,fender}. The radio emission is generally too bright to be produced by thermal electrons in the accretion flow. 

A strong correlation has been found between the radio and X-ray luminosity in the hard state \citep{corbel2003}. This suggests that the X-ray emission is strongly coupled to that of the jets \citep[e.g.,][]{falcke}. Since the jets flow out of the corona and are thus tightly associated to it, a model in which the X-ray emission is mostly from the disk+corona and radio is from the jets is compatible with the observations \citep{narayan}. Other possibilities are discussed, for instance, by \citet{markoff2003}.

In two-temperature magnetized plasmas around black holes it is reasonable to expect significant deviations from purely Maxwellian distributions for the particles. Such deviations are the result of the injection of non-thermal populations of particles. Recently, \citet{vurm} have presented a model for time-dependent non-thermal (purely leptonic) emission from magnetized coronae. \citet{romero03} have studied the case of a steady state magnetized corona with injection of both relativistic electrons and protons. In the present work we extend the latter study to time-dependent injection and calculate, by first time, the neutrino output of a hot magnetized corona around a galactic black hole.     

\section{Static corona model}
 
\subsection{Basic scenario}

The model considered here represents a static corona with a component of relativistic particles that can be removed by diffusion (see \citealt{romero03} for details). In the Bohm regime, the diffusion coefficient is $D(E)=r_{\rm{g}}c/3$, where $r_{\rm{g}}=E/(eB)$ is the gyro-radius of the particles. The diffusion rate is

	\begin{equation}\label{eq:diff}
				t_{\rm{diff}}^{-1}=\frac{2D(E)}{R_{\rm{c}}^2} .
		\end{equation}
		
	Relativistic particles injected in the corona can have a local origin. Strong shocks resulting from fast magnetic reconnection events and converging magnetic mirrors can in principle accelerate particles up to relativistic energies through first-order Fermi mechanism \citep[e.g.,][]{tsuneta,deGouveia}. More complicated processes resulting in second-order Fermi acceleration are plausible in a turbulent corona \citep[e.g.,][]{dermer}. In the present work we assume first-order Fermi acceleration. In such a case, the acceleration rate for a particle of energy $E$ in a magnetic field $B$ is given by

\begin{equation}
	t^{-1}_{\rm{acc}}=\frac{\eta ecB}{E},
	\label{accrate}
\end{equation}
   
\noindent where $\eta\leq1$ is a parameter that characterizes the efficiency of the acceleration. We fix $\eta=10^{-2}$, which describes the efficient acceleration by shocks with $v_{\rm s}\sim 0.1c$ in the Bohm regime.

We consider a two-temperature corona in steady state, with a thermal emission characterized by a power-law with an exponential cutoff at high energies, as observed in several X-ray binaries in the low-hard state \citep[e.g.,][]{romero01}. Since the corona is in steady state, the assumption of equipartition between the different component allows to estimate the mean value of the main parameters in the model. Table (\ref{table}) summarizes these values.

\begin{table}[!h]
\caption{Model parameters}
\begin{tabular}{ll}
\hline
Parameter & Value\\[0.01cm]
\hline
$M_{\rm{BH}}$:  black hole mass [$M_{\odot}$]							& $10$$^{(1)}$					\\[0.01cm]
$R_{\rm{c}}$:   corona radius [cm] 										 	  & $5.2 \times 10^{7}$$^{(1,2)}$ 	\\[0.01cm]
$T_{e}$:        electron temperature [K] 								&	$10^9$								\\[0.01cm]
$T_{i}$:        ion temperature [K] 											&	$10^{12}$ 						\\[0.01cm]
$E_{\rm{c}}$:   X-ray spectrum cutoff [keV]							& $150$       					\\[0.01cm]
$\alpha$: 			X-ray spectrum power-law index    				& $1.6$									\\[0.01cm]
$\eta$: 				acceleration efficiency 									& $10^{-2}$							\\[0.01cm]
$B_{\rm{c}}$: 	magnetic field [G]				 								& $5.7 \times 10^5$				\\[0.01cm]
$n_{i},n_{e}$:   plasma density [cm$^{-3}$] 								& $6.2 \times 10^{13}$	\\[0.01cm]
$a$: 						hadron-to-lepton energy ratio 						& $100$	      					\\[0.01cm]
$kT$:						disk characteristic temperature [keV] 	  & $0.1$									\\[0.01cm]

\hline\\[0.005cm]
\multicolumn{2}{l}{
$^{(1)}$ Typical value for Cygnus X-1 in the low-hard state }				  \\[0.01cm]
\multicolumn{2}{l}{
$^{(2)}$ $35 R_{\rm{G}}$, $ R_{\rm{G}}=\frac{GM}{c^{2}}$.} 			 																				\\[0.01cm]
\end{tabular}	
  \label{table}
\end{table} 

\begin{figure*}[!h]
\centering
\subfigure[Electron losses.]{\label{fig:perdidas:a}\includegraphics[width=0.45\textwidth,keepaspectratio]{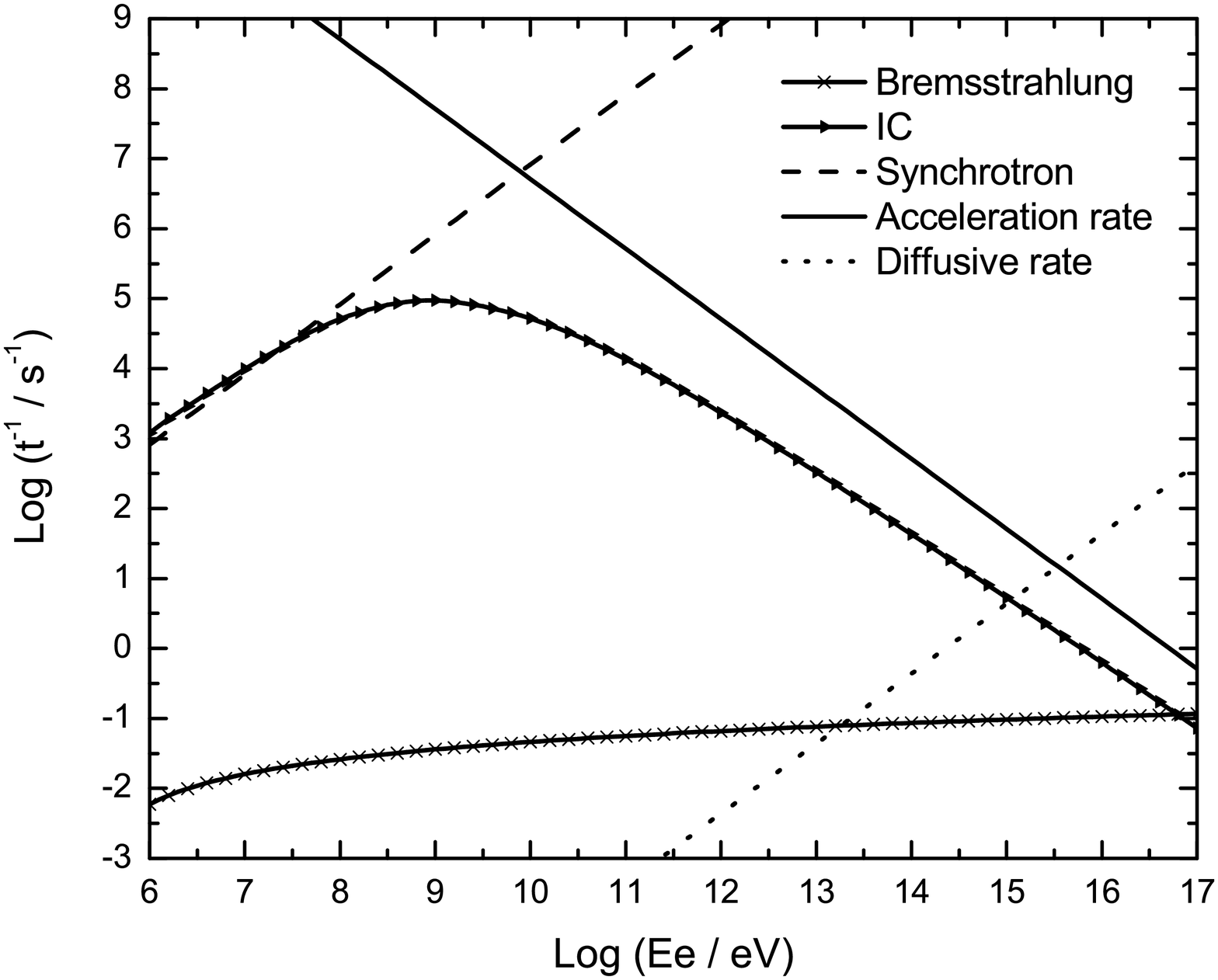}} \hspace{20pt} 
\subfigure[Proton losses.]{\label{fig:perdidas:b}\includegraphics[width=0.45\textwidth,keepaspectratio]{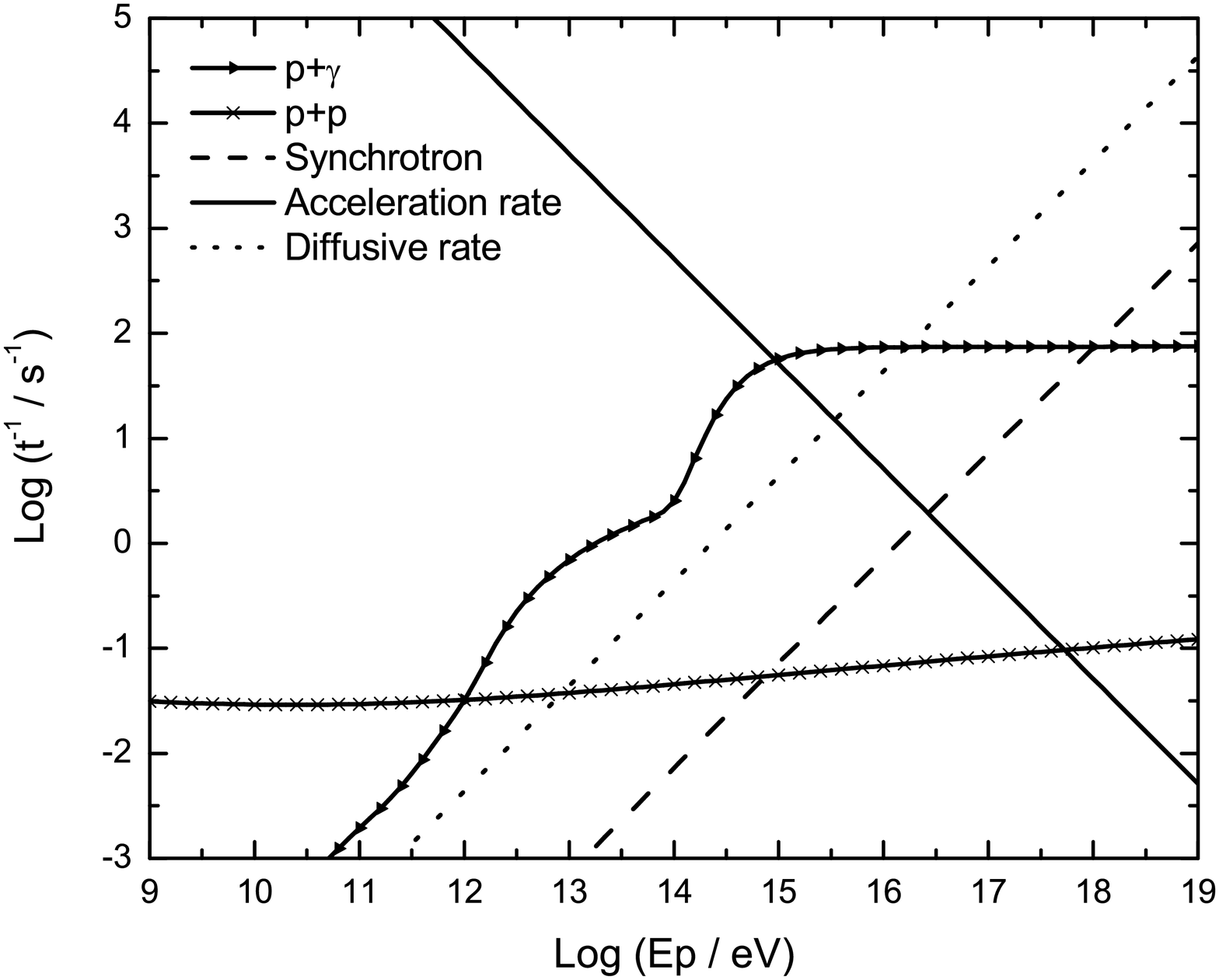}} \hfill \\ 
\subfigure[Pion losses.]{\label{fig:perdidas:c}\includegraphics[width=0.45\textwidth, keepaspectratio]{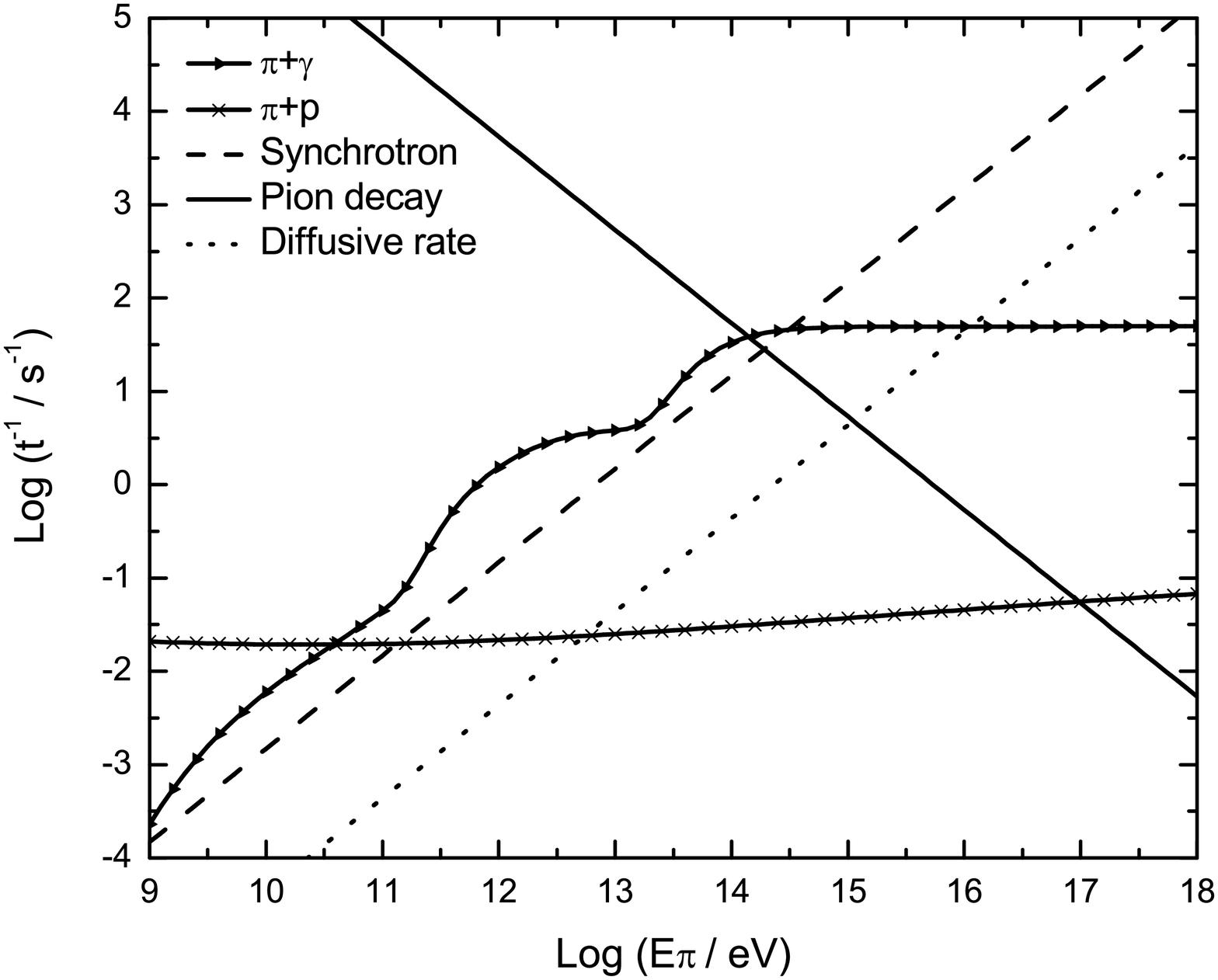}}  \hspace{20pt}
\subfigure[Muon losses.]{\label{fig:perdidas:d}\includegraphics[width=0.45\textwidth, keepaspectratio]{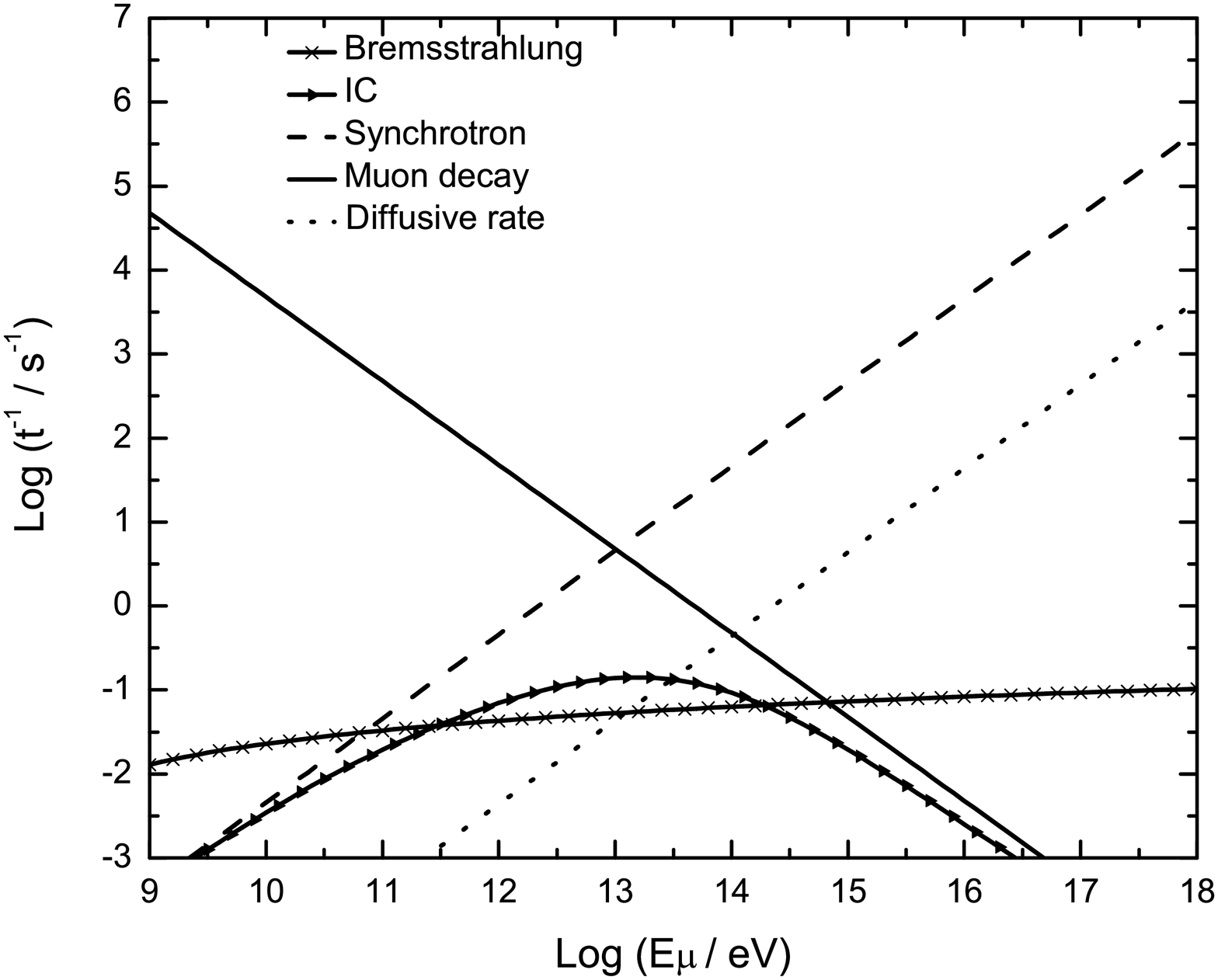}} \hfill
\caption{Radiative losses in a corona characterized by the parameters of Table \ref{table}.}
\label{fig:perdidas}
\end{figure*}

In a corona characterized by such parameters, it is expected that relativistic electrons and muons lose energy mainly because of synchrotron radiation and inverse Compton scattering, whereas for protons and charged pions the relevant cooling processes are synchrotron radiation, inellastic proton-proton collisions, and photomeson production. Figure (\ref{fig:perdidas}) shows the cooling time for the different radiative processes under the conditions of Table (\ref{table}).

\subsection{Spectral energy distributions}

We study the effect of the injection of a power-law distribution of relativistic electrons and protons. The main products of hadronic interactions are charged pions, which quickly decay producing muons and neutrinos. Neutral pions yield gamma-rays, that are a source of secondary pairs. Therefore, we also include the effect of all these secondary particles in our treatment.

\begin{figure}[!h]
\centering
\includegraphics[clip,width=0.6\textwidth, keepaspectratio]{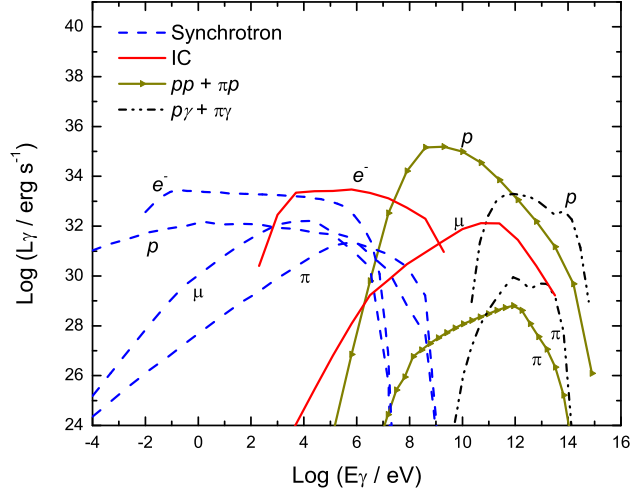}
\caption{Non-thermal contributions to the total luminosity.}
\label{fig:SEDs}
\end{figure}

We solve the transport equation in steady state obtaining particle distributions for the different species. Then, the spectral energy distributions (SEDs) of all radiative processes are estimated. The results are shown in Fig. (\ref{fig:SEDs}).

\begin{figure}[!h]
\centering
\includegraphics[clip,width=0.6\textwidth, keepaspectratio]{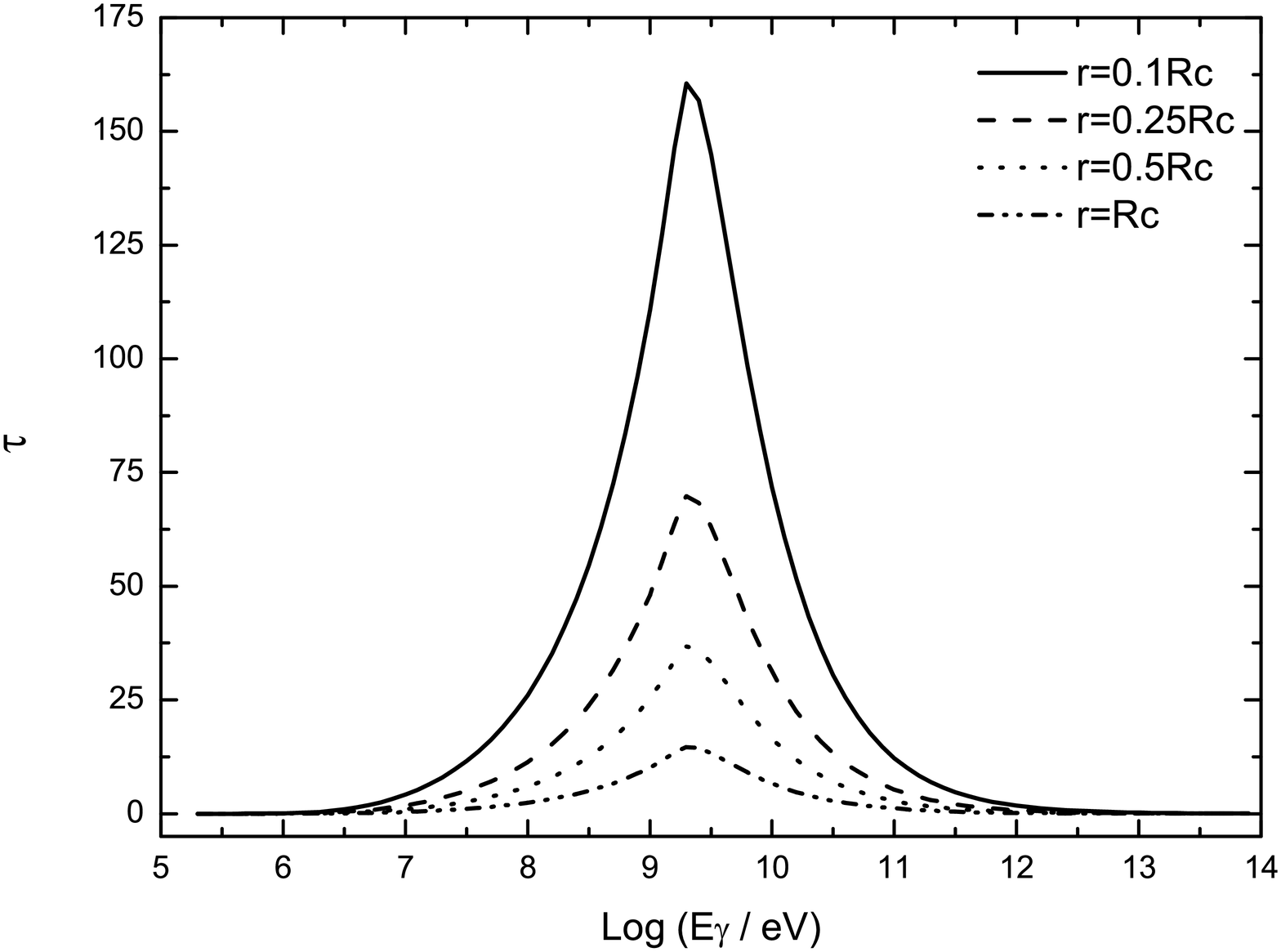}
\caption{Internal absorption due to photon-photon pair production.}
\label{fig:opacity}
\end{figure}

Gamma-rays produced in the corona can be absorbed by different mechanisms. The most relevant one is photon-photon annihilation. The absorption can be quantified by the absorption coefficient or opacity $\tau$. In Fig. (\ref{fig:opacity}) we show the opacity due to the interaction between gamma-rays and thermal X-ray photons from the corona, which are by far the dominant electromagnetic component at low energies. 

\begin{figure}[!h]
\centering
\includegraphics[clip,width=0.6\textwidth, keepaspectratio]{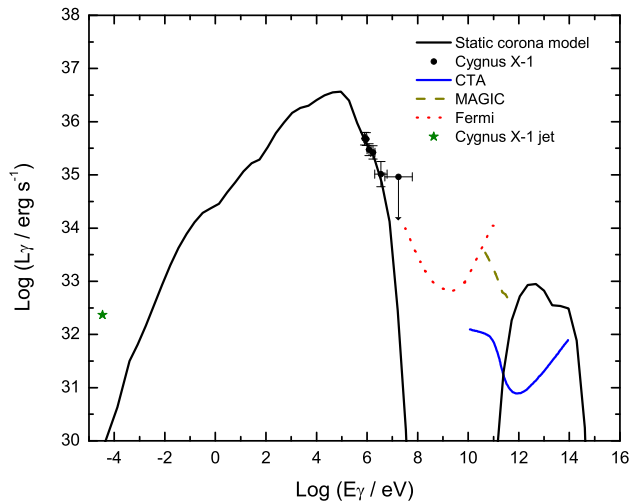}
\caption{Spectral energy distribution obtained with the steady state model of a static corona. The prediction fits the observation made by COMPTEL and {\it INTEGRAL} of Cygnus X-1 (\citealt{McConnell,cadolle}). The non-thermal radio emission from the jet is also shown \citep{stirling}.}
\label{fig:ajuste}
\end{figure}

Because of the high values of the opacity, it is expected a large number of secondary pairs. The effects of internal absorption and the radiation emitted by secondary pairs are also include in the final SED, which is shown in Fig. (\ref{fig:ajuste}). This figure also shows the spectrum of the well-known source Cygnus X-1, detected by COMPTEL \citep{McConnell} and {\it INTEGRAL} \citep{cadolle}, and the radio emission from the jet. As it can be seen, the emission of the corona at low energy is negligible compared to that of the jet, in accordance with the idea that both components are present in the \textit{low-hard} state.

\subsection{Mass Scaling}

Currently, the best evidence for the existence of a common mechanism operating in both X-ray binaries and Active Galactic Nuclei (AGNs) is given by the detection of steady jets in accretion regimes with low rates on all scales \citep{markoff2005}. It is thought that most black holes spend a significant amount of time in this low-luminosity regime on their way in and out of the quiescent ground state. For galactic black holes this corresponds to the hard state. In AGNs this would correspond to the class of low-luminosity AGNs (LLAGNs; e.g., \citealt{ho}) which includes most nearby AGNs such as M81, NGC 4258 and M31. 

The latest results strongly support that accretion at low rates is similar across the entire range of black hole masses. It seems that the same physical model, in which all parameters were expressed in mass-scaling units (such as Eddington luminosity and $r_{\rm{g}}$), should be able to describe the spectra of either a hard state XRB or a LLAGN, with similar internal parameters. 

In \citet{markoff2010} is shown that the same outflow-dominated (including an outflowing corona as the jet base) model can provide statistically good descriptions of broadband data from hard state XRBs as well as LLAGNs, with similar ranges in free parameters. These results agree with the idea that the physics of at least weakly accreting black holes scales predictably with mass. The data are consistent, across the mass scale, with a compact outflowing corona directly feeding a continuous, steady, weakly accelerated jet.

All this suggests that the results presented on this paper can be scaled to AGN coronae as well.

\section{Flare model}

\subsection{Particle injection}

\begin{figure}[!h]
\centering
\includegraphics[clip,width=0.5\textwidth, keepaspectratio]{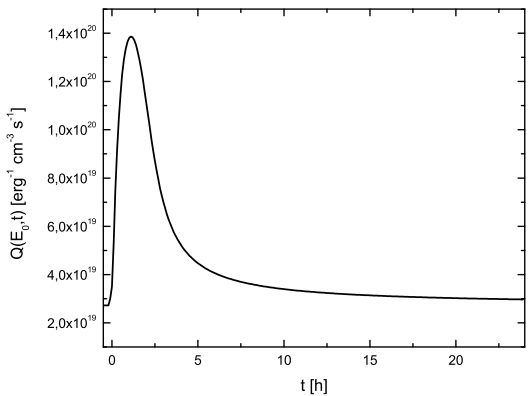}
\caption{Electron injection as a time function at $E_{0}=10^7$ eV.}
\label{fig:injection}
\end{figure}

The temporal dependence of the assumed particle injection is characterized by a FRED (Fast Rise and Exponential Decay) behavior, whereas the energy dependence is a power-law. The transient injection  can be represented by

\begin{equation}
Q(E,t) = Q_{0} E^{-\alpha} e^{-E/E_{\rm{max}}} (1-e^{t/\tau_{\rm{rise}}} ) \left[ \frac{\pi}{2}- \arctan \Big( \frac{t-\tau_{\rm{plat}}}{\tau_{\rm{dec}}} \Big) \right],
\end{equation}

\noindent where $\tau_{\rm{rise}} = 30$ min, $\tau_{\rm{dec}} = 1$ h and $\tau_{\rm{plat}} = 2$ h. Figure (\ref{fig:injection}) shows the primary electron injection at $E=10^7$ eV. A similar injection takes place for protons.

\noindent The power-law has the standard index of $\alpha=2.2$. The normalization constant $Q_{0}$ can be obtained from the total power injected in relativistic protons and electrons, $L_{\rm{rel}}=L_{p}+L_{e}$. This power is assumed to be a fraction of the luminosity of the corona, $L_{\rm{rel}}=q_{\rm rel} L_{\rm{c}}$. In the steady state the best fit to the observations is obtained with $q_{\rm rel}=0.2$. During the flare the number of relativistic particles increases. In our model, the power injected in the flare doubles that of the steady state. In solar flares however, the non-thermal injection can be much higher \citep{lin} suggesting that stronger flares can occur in XRB coronae due to massive magnetic reconnection.

The way in which energy is divided between hadrons and leptons is unknown, but different scenarios can be taken into account by setting $L_{p}=aL_{e}$. We consider a model with $a=100$ (proton-dominated scenario, as for galactic cosmic rays).

\subsection{Transport equation}

The particle distributions $N(E,t)$ can be derived from the solution to the transport equation \citep{ginzburg}:

\begin{equation}\label{eq:transporte}
		\frac{\partial N(E,t)}{\partial t} +  \frac{\partial }{\partial E} \Big( b(E) N(E,t) \Big)+ \frac{N(E,t)}{t_{\rm{esc}}}=Q(E,t) ,
	\end{equation} 

\noindent where $b(E)= \frac{dE}{dt} \Big | _{\rm{loss}}$. The corresponding solution is:

\begin{equation}
			N(E,t)= \frac{1}{\Big | b(E) \Big | } \int_{E}^{E_{\rm{eff}}} {Q(E',t-\tau)e^{-\tau (E,E')/t_{\rm{esc}}}dE'} ,
		\end{equation}

\noindent where $E_{\rm{eff}}$ is implicity defined by

    \begin{equation} \label{eq:tau}
			t = \int_{E}^{E_{\rm{eff}}} { \frac{ dE'}{ \Big |b(E') \Big | }} ,
		\end{equation}
		
\noindent and
	
		\begin{equation} \label{eq:tau}
			\tau (E,E')= \int_{E}^{E'} { \frac{dE''}{ \Big |b(E'') \Big | }} .
		\end{equation}

\begin{figure*}[!ht]
\centering
\includegraphics[width=0.7\textwidth,keepaspectratio]{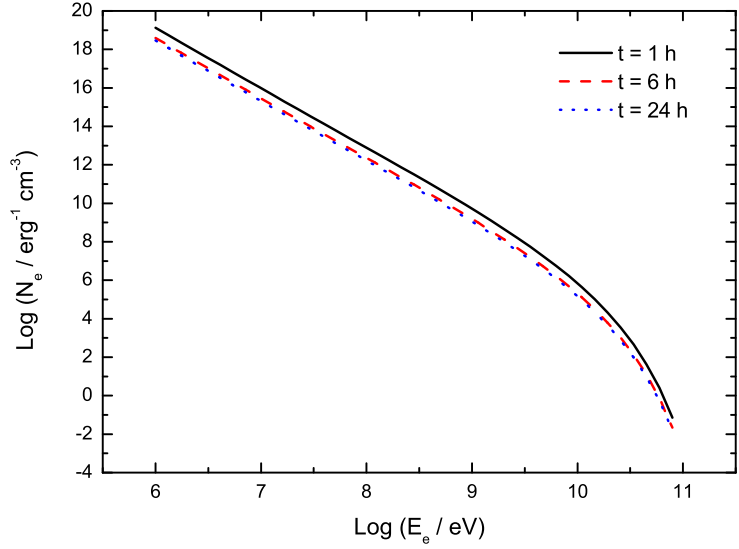}
\caption{Evolution in time of electron distribution.}
\label{fig:electron}
\end{figure*}

The primary electron distribution is shown in Fig. (\ref{fig:electron}). Similar distributions where calculated for all types of particles: $p$, $\pi^{\pm}$, $\pi^{0}$, $\mu^{\pm}$ and secondary $e^{\pm}$.

\subsection{Spectral energy distribution}

\begin{figure*}[!ht]
\centering
\subfigure[$E=1$ MeV.]{\label{fig:1MeV:a}\includegraphics[width=0.45\textwidth,keepaspectratio]{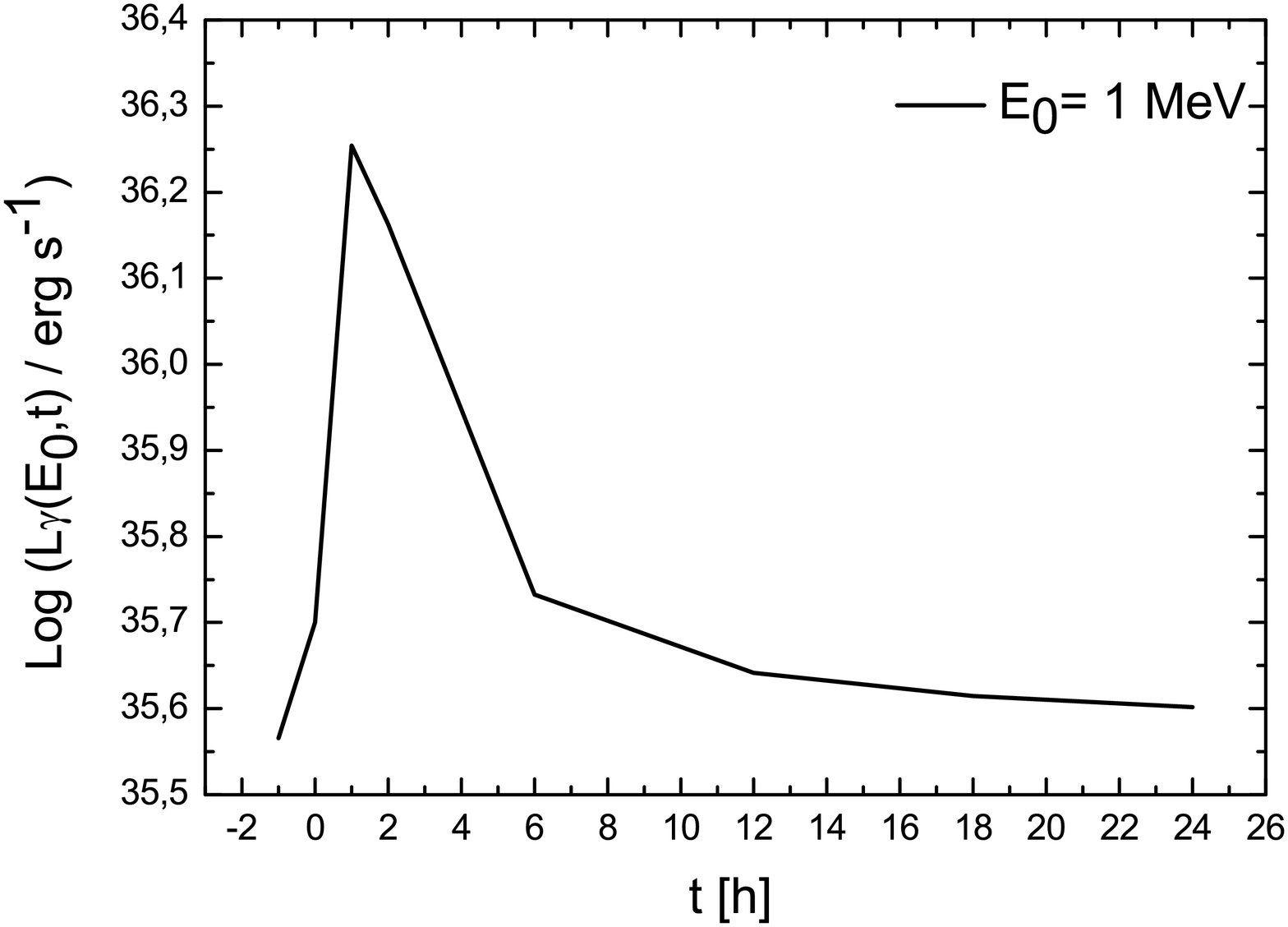}} \hspace{20pt} 
\subfigure[$E=1$ TeV.]{\label{fig:1TeV:b}\includegraphics[width=0.45\textwidth,keepaspectratio]{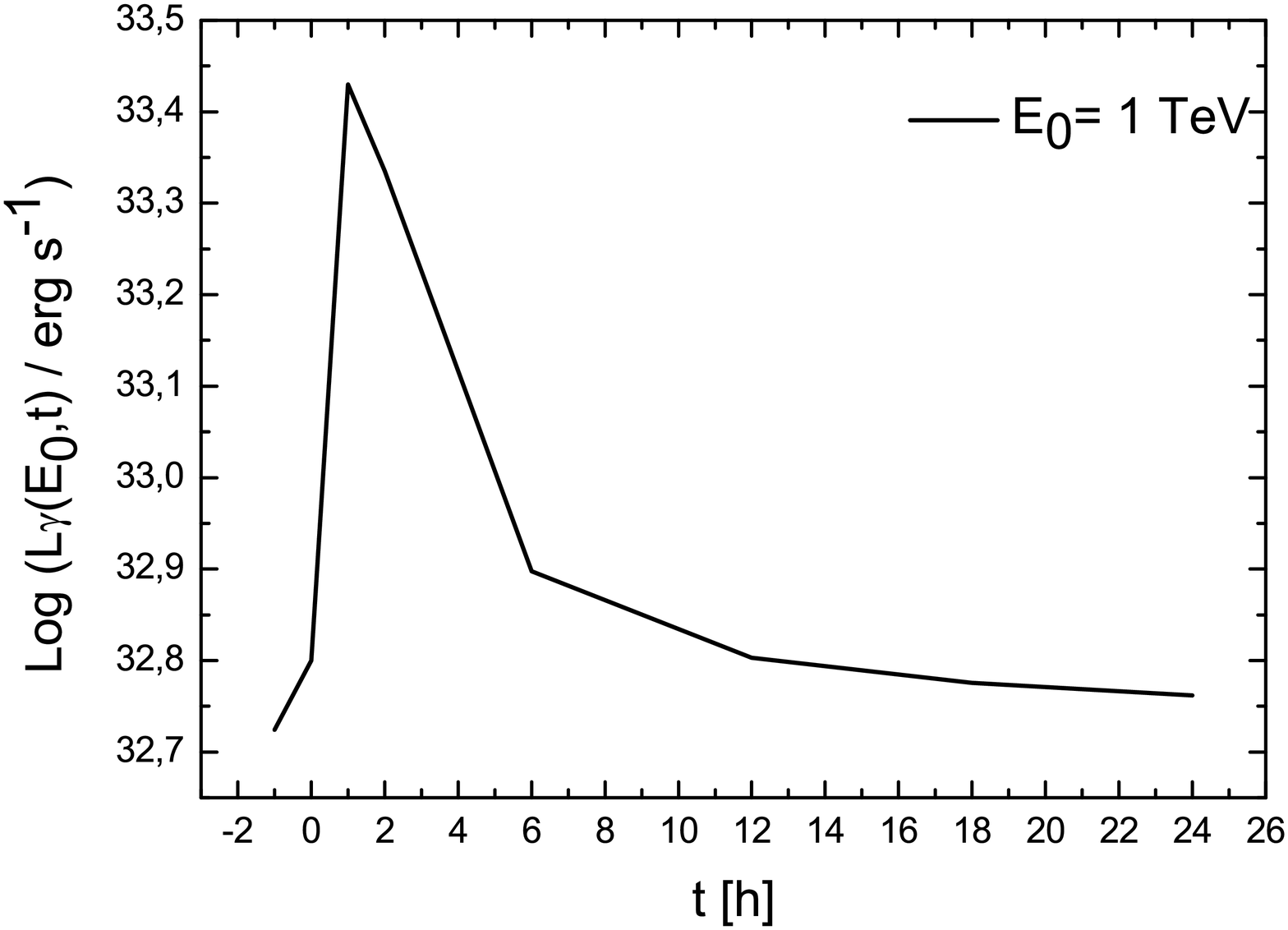}} \hfill \\ 
\caption{SEDs evolution at different energies. The sensitivity of CTA at $E=1$ TeV is $2.62 \times 10^{-14}$ TeV cm$^{-2}$ s$^{-1}$, which corresponds to a luminosity of $7.8 \times 10^{30}$ erg s$^{-1}$ at a distance of 2 kpc. Thus the emission might be detectable.}
\label{fig:SEDevolution}
\end{figure*}

Figure (\ref{fig:SEDevolution}) shows the evolution of the total non-thermal luminosity absorbed by photon-photon pair production at different energies. The emission at $E=1$ MeV is dominated by the thermal contribution from the corona.

\subsection{Neutrino emission}

The secondary particles produced by hadronic and photohadronic interactions decay as:

\begin{equation}
		 \pi^{\pm} \rightarrow \mu^{\pm} + \nu_{\mu}(\overline{\nu}_{\mu}) \textrm{,}
		\end{equation}
	
	\begin{equation}
		 \mu^{\pm} \rightarrow e^{\pm} + \overline{\nu}_{\mu}(\nu_{\mu}) + \nu_{e}(\overline{\nu}_{e}) .
		\end{equation}
		
\noindent We consider neutrino production by these two channels. Thus, the total emissivity of neutrinos is \citep{reynoso}:

\begin{equation}
Q_{\nu}(E,t) = Q_{\pi \rightarrow \nu}(E,t) + Q_{\mu \rightarrow \nu}(E,t) ,
\end{equation}

\noindent where

	\begin{equation}
		Q_{\pi \rightarrow \nu}(E,t) = \int^{E^{\rm{max}}}_{E} dE_{\pi} t^{-1}_{\pi,\rm{ dec}}(E_{\pi})N_{\pi}(E_{\pi},t) \frac{\Theta(1-r_{\pi}-x) }{E_{\pi}(1-r_{\pi})},
	\end{equation}
	
\noindent with $x=E/E_{\pi}$, and

	\begin{equation}
		Q_{\mu \rightarrow \nu}(E,t) = \sum^4_{i=1} \int^{E^{\rm{max}}}_{E} \frac{dE_{\mu}}{E_{\mu}} t^{-1}_{\mu,\rm{ dec}}(E_{\mu})N_{\mu_{i}}(E_{\mu},t)
		 \left[ \frac{5}{3}-3x^2+ \frac{4}{3}x^3 \right].
	\end{equation}
	
\noindent In this latter expression, $x=E/E_{\mu}$, $\mu_{\{1,2\}}=\mu_{\rm{L}}^{\{-,+\}}$, and $\mu_{\{3,4\}}=\mu_{\rm{R}}^{\{-,+\}}$ .

The total cosmic ray energy density in a source at a distance $d$ to display detectable neutrino emission is $L_{p}/4\pi d^2 c$. Assuming the distance to Cygnus X-1 ($d \approx 2$ kpc), we obtain a cosmic ray energy density of $2.24\times10^{-7}$ eV cm$^{-3}$ at the peak of the emission.

The differential flux of neutrinos arriving at the Earth can be obtained as:

\begin{equation}
\frac{d \Phi_{\nu}}{dE} = \frac{1}{4\pi d^2} \int_{\rm{V}}{d^3r Q_{\nu}(E,t)} .
\end{equation}

\begin{figure}[!h]
\centering
\includegraphics[clip,width=0.7\textwidth, keepaspectratio]{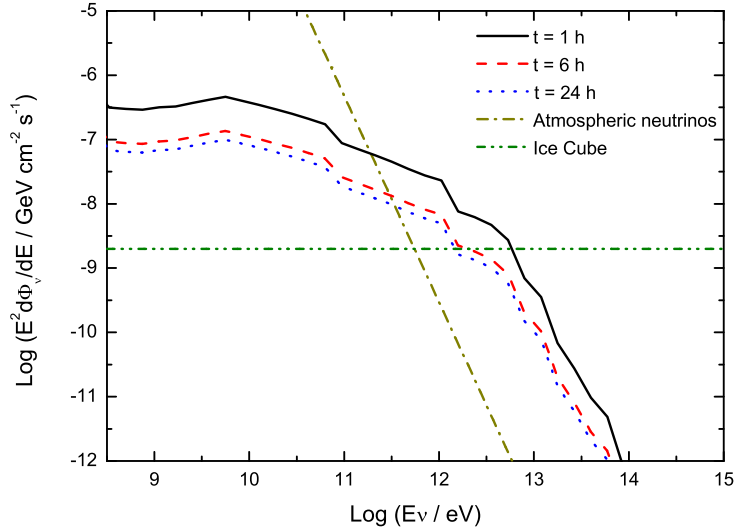}
\caption{Evolution of the neutrino flux, atmospheric neutrino flux and Ice Cube sensitivity.}
\label{fig:neutrinoFlux}
\end{figure}

This quantity, weighted by the squared energy, is shown in Fig. (\ref{fig:neutrinoFlux}). Also the IceCube sensitivity for 1 year of operation is shown in the figure. Assuming that the duty cycle of flares in galactic black holes is around $10$ \%, the Ice Cube detector will be able of detecting neutrinos from a source at $\sim 2$ kpc after $10$ years of observations.

\section{Summary and conclusions}

According to our results, the accumulated signal due to neutrino bursts in black hole coronae might be detectable for sources within a few kpc. The electromagnetic part of the flare can also be detectable at some energies, depending of the optical depth for photon annihilation. It should be remarked that in some cases the source can be `dark' at gamma rays and nonetheless a significant neutrino emitter (e.g. Cygnus X-3, see \citealt{cerutti} for estimates of the $\gamma-\gamma$ annihilation).

In a future work, we will take into account the effects of absorption due to the presence of the anisotropic photon field of the donor star. This absorption depends on the position in the orbit of the compact object \citep{romero02, vieyro}. Another effect that will be studied, is the electromagnetic flare modulation produced by the star. This effect is relevant only in sources where the duration of the flare is comparable with the period of the compact object. Particular emphasis will be put on the source Cygnus X-3, which presents a short period of less than $5$ hours.

\section{Acknowledgments}
This research was supported by the Argentine Agencies CONICET and ANPCyT through grants PIP 0078 and PICT-2007-00848 BID 1728/OC-AR and by the Ministerio de Educaci\'{o}n y Ciencia (Spain) under grant AYA2010-21782-C03-01, FEDER funds.





\begin{thebibliography}{99}
\expandafter\ifx\csname natexlab\endcsname\relax\def\natexlab#1{#1}\fi

\bibitem[{{Blandford} \& {Eichler}(1987){Blandford} \& {Eichler}}]{blandford02}
{Blandford}, R.~D., {Eichler}, D. Particle acceleration at astrophysical shocks: A theory of cosmic ray origin, Phys. Rep., 154, p. 1-75, 1987.

\bibitem[{{Cadolle Bel} {et~al.}(2006){Cadolle Bel} {et~al.}}]{cadolle}
{Cadolle Bel}, M., {et~al.} The broad-band spectrum of Cygnus X-1 measured by INTEGRAL, A\&A, 446, p. 591-602, 2006.

\bibitem[{{Cerutti} {et~al.}(2011){Cerutti} {et~al.}}]{cerutti}
{Cerutti} B., {et~al.} Absorption of high-energy gamma rays in Cygnus X-3, A\&A, in press [arXiv:1103.3875], 2011.

\bibitem[{{Corbel} {et~al.}(2000){Corbel} {et~al.}}]{corbel2000}
{Corbel} S., {et~al.} Coupling of the X-ray and radio emission in the black hole candidate and compact jet source GX 339-4, A\&A, 359, p. 251-268, 2000.

\bibitem[{{Corbel} {et~al.}(2003){Corbel} {et~al.}}]{corbel2003}
{Corbel} S., {et~al.} Radio/X-ray correlation in the low/hard state of GX 339-4, A\&A, 400, p. 1007-1012, 2003.

\bibitem[{{Dermer} {et~al.}(1996){Dermer} {et~al.}}]{dermer} 
{Dermer}, C., {Miller}, J.~A., {Li}, H. Stochastic Particle Acceleration near Accreting Black Holes, ,ApJ, v.456, p. 106-118, 1996.

\bibitem[{{de Gouveia Dal Pino} {et~al.}(2010){de Gouveia Dal Pino} {et~al.}}]{deGouveia} 
{de Gouveia Dal Pino}, E.~M., {Piovezan}, P.~P., {Kadowaki}, L.~H.~S. The role of magnetic reconnection on jet/accretion disk systems, A\&A, 518, A5, p. 1-9, 2010.

\bibitem[{{Falcke} {et~al.}(2004){Falcke}, {Koerding} \& {Markoff}}]{falcke}
{Falcke}, H., {Koerding}, E., {Markoff}, S. A scheme to unify low-power accreting black holes. Jet-dominated accretion flows and the radio/X-ray correlation, A\&A, 414, p. 895-903, 2004.

\bibitem[{{Fender} \& {Belloni}(2004){Fender} \& {Belloni}}]{fender}
{Fender}, R., {Belloni} T. GRS 1915+105 and the Disc-Jet Coupling in Accreting Black Hole Systems, A\&A, 42, p. 317-364, 2004.

\bibitem[{{Ginzburg} \& {Syrovatskii}(1964){Ginzburg} \& {Syrovatskii}}]{ginzburg}
{Ginzburg}, V.~L., {Syrovatskii} S.~I. \emph{The Origin of Cosmic Rays}, Macmillan, New York, 1964.

\bibitem[{{Ho}(2005){Ho}}]{ho}
{Ho}, L.~C. The Spectral Energy Distributions of Low-Luminosity Active Galactic Nuclei, ApJ, 516, p. 672-682, 1999.

\bibitem[{{Lin}(2008){Lin}}]{lin}
{Lin}, R.~P. Particle Acceleration by the Sun, Particle Acceleration and Transport in the Heliosphere and beyond: 7th Annual International Astrophysics Conference. AIP Conference Proceedings, 1039, p. 52-62, 2008.

\bibitem[{{Markoff} {et~al.}(2003){Markoff} {et~al.}}]{markoff2003}
{Markoff}, S., {et~al.} Exploring the role of jets in the radio/X-ray correlations of GX 339-4, A\&A, 397, p. 645-658, 2003.

\bibitem[{{Markoff}(2005){Markoff}}]{markoff2005}
{Markoff}, S. Accretion and Jets in Microquasars and Active Galactic Nuclei, ASP Conference Series, Vol. 352, p. 129-144, 2005.

\bibitem[{{Markoff}(2010){Markoff}}]{markoff2010}
{Markoff}, S. From Multiwavelength to Mass Scaling: Accretion and Ejection in Microquasars and AGN, LNP, Vol. 794, p. 143-174, 2010.

\bibitem[{{McConnell} {et~al.}(2000){McConnell} {et~al.}}]{McConnell}
{McConnell}, M.~L., {et~al.} A High-Sensitivity Measurement of the MeV Gamma-Ray Spectrum of Cygnus X-1, ApJ, 543, p. 928-937, 2000.

\bibitem[{{McConnell} {et~al.}(2002){McConnell} {et~al.}}]{McConnell2002}
{McConnell}, M.~L., {et~al.} The Soft Gamma-Ray Spectral Variability of Cygnus X-1, ApJ, 572, p. 984-995, 2002.

\bibitem[{{Narayan} \& {McClintock}(2008){Narayan} \& {McClintock}}]{narayan}
{Narayan}, R., {McClintock}, J.~E. Advection-dominated accretion and the black hole event horizon, New Astronomy Reviews, 51, Issue 10-12, p. 733-751, 2008.

\bibitem[{{Poutanen} (1998){Poutanen}}]{poutanen03}
{Poutanen}, J. Accretion disc-corona models and X/$\gamma$-ray spectra of accreting black holes, in M. A. Abramowicz,  G. Bjornsson, and J. E. Pringle (eds), \emph{Theory of Black Hole Accretion Disks}, Cambridge University Press, p.100-122, 1998.

\bibitem[{{Reynoso} \& {Romero}(2009){Reynoso} \& {Romero}}]{reynoso}
{Reynoso}, M., {Romero}, G.~E. Magnetic field effects on neutrino production in microquasars, A\&A, 493, 1, p. 1-11, 2009.

\bibitem[{{Romero} {et~al.}(2002){Romero}, {Kaufman Bernad\'o} \& {Mirabel}}]{romero01}
{Romero}, G.~E., {Kaufman Bernad\'o}, M.~M., {Mirabel}, I.~F. Recurrent microblazar activity in Cygnus X-1?, A\&A, 393, p. L61-L64, 2002.

\bibitem[{{Romero} {et~al.}(2010){Romero}, {del Valle} \& {Orellana}}]{romero02}
{Romero}, G.~E., {del Valle}, M.~V., {Orellana}, M. Gamma-ray absorption and the origin of the gamma-ray flare in Cygnus X-1, A\&A, 518, A12, p. 1-9, 2010.

\bibitem[{{Romero} {et~al.}(2010){Romero}, {Vieyro} \& {Vila}}]{romero03}
{Romero}, G.~E., {Vieyro}, F.~L., {Vila}, G.~S. Non-thermal processes around accreting galactic black holes, A\&A, 519, A109, p. 1-11, 2010.

\bibitem[{{Romero} \& {Vieyro}(2011){Romero} \& {Vieyro}}]{vieyro}
{Romero}, G.~E., {Vieyro}, F.~L. Gamma-ray flares from black hole coronae, Proceedings of the 25th Texas Symposium on Relativistic Astrophysics, in press, 2011.

\bibitem[{{Shakura} \& {Sunyaev}(1973){Shakura} \& {Sunyaev}}]{shakura}
{Shakura}, N.~I., {Sunyaev}, R.~A. Black holes in binary systems. Observational appearance, A\&A, 24, p. 337-355, 1973.

\bibitem[{{Shapiro} {et~al.}(1976){Shapiro}, {Lightman} \& {Eardley}}]{shapiro}
{Shapiro}, S.~L., {Lightman}, A.~P., {Eardley}, D.~M. Black holes in X-ray binaries - Marginal existence and rotation reversals of accretion disks, ApJ, v. 204, pt. 1, p. 187-199, 1976.

\bibitem[{{Stirling} {et~al.}(2001){Stirling} {et~al.}}]{stirling}
{Stirling}, A.~M., et al. A relativistic jet from Cygnus X-1 in the low/hard X-ray state, MNRAS, 327, p. 1273-1278, 2001.

\bibitem[{{Tsuneta} \& {Naito}(1998){Tsuneta} \& {Naito}}]{tsuneta}
{Tsuneta}, S., {Naito} T. Fermi Acceleration at the Fast Shock in a Solar Flare and the Impulsive Loop-Top Hard X-Ray Source, ApJ, 495, p. L67-L70, 1998.

\bibitem[{{Vurm} \& {Poutanen}(2009){Vurm} \& {Poutanen}}]{vurm}
{Vurm}, I., {Poutanen}, J. Time-Dependent Modeling of Radiative Processes in Hot Magnetized Plasmas, ApJ, 698, p. 293-316, 2009.


\end{thebibliography}

\end{document}